# Low temperature charge ordering versus grain boundary effects in polycrystalline $La_{1-x}Ca_xMnO_3$ manganites: comment on paper "Magnetoresistance in $La_{1-x}Ca_xMnO_3$ ($0 \leq x < 0.4$)"

E. Rozenberg*

*Department of Physics, Ben-Gurion University of the Negev, POB 653, 84105 Beer-Sheva, Israel*

**Abstract**

In recent paper G. Li et al, (Solid State Communications, 128 (2003) 171) reported on magnetic properties and magnetoresistance (MR) in $La_{1-x}Ca_xMnO_3$ ($0 \leq x < 0.4$) system. Authors claimed that the changes in MR versus temperature ($T$) dependences, e.g., broadening of its 'profile' due to increasing of MR values at low $T$ (below the Curie point - $T_C$) for $x > 0.25$ are induced by increasing content of short range charge ordered clusters in ferromagnetic matrix of such samples. It is shown that (i) the samples used by authors are badly characterized and (ii) the above interpretation of the effects observed is very questionable. The increasing of MR values at $T \ll T_C$ due to the well-known grain boundary effects in polycrystalline manganites seems to be the simplest and natural explanation.

* Corresponding author. Tel.: +972-8-6472456; fax: +972-8-6472903.

*E-mail address*: evgenyr@bgumail.bgu.ac.il

After recent "rediscovery" of colossal magnetoresistance and related phenomena in mixed valence doped manganites $Ln_{1-x}A_xMnO_3$ (Ln = La, Pr, Nd etc. and A = Ca, Sr, Ba etc.) the studying of its properties (closely connected magnetic and conductive ones in particular) develop intensively - see, e.g., [1-4]. La-manganite doped with Ca and having chemical formula $La_{1-x}Ca_xMnO_3$ (LCMO) is among the most popular systems under investigation. One may roughly estimate the number of papers published during last decade, in which the physical properties of LCMO were studied, as being about some hundreds - see [1-4] and references therein. Thus, it is obvious that any new results regarding the magnetic and magnetotransport properties of LCMO must be compared carefully with the known ones for this peculiar system. Let us discuss in this context the paper [5] commented upon.

(i) *The samples studied* in [5]. As noted by G. Li et al. [5] their parent ceramic $LaMnO_3$ (LMO) compound ($x = 0$) has antiferromagnetic (AFM) ground state below the Neél temperature $T_N \sim 160$ K - see Fig. 2 in [5]. This value is considerably higher than $T_N \sim 140$ K reported for LMO by numerous researchers [3,4] and is out of the normal experimental error in temperature measurements. Very recent data of Ref. [6] evidence that the vacancies in La-Mn sites may result in the transition at $T \sim 165$ K for LMO. But, in this case the Curie point ($T_C$) and ferromagnetic (FM) ground state are observed [6]. So, the above contradiction must be realized and clarified by authors (reviews [3,4] were cited by G. Li et al.). But it was not done in [5] at all. Thus, starting from such non-trivial LMO, G. Li et al. prepared LCMO ceramics with $0 \leq x < 0.4$ by the conventional solid-state reaction method. The following important remarks regarding these samples must be addressed.

Firstly, the experimental error in definition of Ca content ($x$) is not less than about 0.03. It may be concluded using dependence of resistivity ($\rho$) vs. $T$ for $x = 0.2$ sample in Fig. 3 from [5]. Namely, this ρ vs. *T* curve demonstrate only a sharp maximum at some temperature. I.e., the typical transition to FM metallic-like (FMM) ground state at $T < T_C$ is observed [1-4]. At the same time, neutron diffraction experiments [7] carried out on perfect single crystals of LCMO show that in this system transition to FMM ground state occurs above the critical concentration $x_c = 0.22$. At $x < x_c$ beginning from $x \approx 0.15$ mixed FMM+FMI (FM metallic and insulating) state exists below $T_C$, which results in local maximum of ρ vs. *T* at some *T* close to $T_C$ and the following low temperature upturn of ρ - see, for example, Refs. [8] (cited by authors) and [9,10]. Comparing all these data, one can easily conclude that the sample, denoted by authors of [5] as LCMO with $x = 0.2$, must in reality have Ca-doping level $x > x_c = 0.22$.

Secondly, it is directly seen from Fig. 3 in [5] that well known grain boundary (GB) effects on ρ vs. *T* curves (see, e.g., [11-13]) become more and more pronounced for the samples with Ca doping $x \geq 0.3$. Namely, for samples with $x = 0.3$ and 0.35 an additional 'shoulders' are observed at temperatures below $T^*$, at which sharp maxima for ρ vs. *T* exist. Further, for $x = 0.37$ and 0.39 above noted sharp maxima transform to broad ones at $T_{max} < T_C$. Except of this, in $x = 0.39$ sample under magnetic field $H = 0$ the two maxima on ρ vs. *T* curve exist: broad and sharp ones at $T_{max}$ and $T^*$, respectively. Note also that shallow minima at $T < 50$ K are clearly observed for all $x = 0.3 - 0.39$ samples - Fig. 3 in [5]. Direct connection of such drastic changes in shape of ρ vs. *T* curves with GB effects (spin-dependent tunneling of charge carriers through GB region etc.) is evidently and impressively illustrates, for example, in Fig. 2 from [13] for an epitaxial bicrystal film of LCMO ($x = 0.33$) with an artificially created single grain boundary.

The next strong argument in favor of GB-originated nature of broad maxima at $T_{max} < T_C$ and shallow minima at low $T$ is its field dependences for $x \geq 0.3$ samples [5]. Namely, only the 'heights' of the above maxima were decreased under external magnetic field, whereas $T_{max}$ values are weakly changed by $H$, while the sharp maxima of ρ near $T_C$ is notably shifted to higher $T$ and suppressed by the same $H$ - compare ρ ($T, H$) curves for $x = 0.2; 0.25; 0.3$ and $x = 0.37; 0.39$ samples on Fig. 3 in [5]. At the same time, the shallow minima at low $T$ are suppressed by external $H$. All these observations agree pretty well with the field dependences of GB-originated features on ρ vs. $T$ curves for different polycrystalline manganites [12, 17, 18].

(ii) *The interpretation of the results* presented in [5]. The main idea addressed by G. Li et al to explain the increasing of MR values at low $T$ observed experimentally in $x > 0.25$ samples (Fig. 4 from [5]) is the following. The short-range charge ordering (CO) correlations are almost absent in their LCMO samples with $x = 0.2 - 0.25$ at $T < T_C$, while such CO/AFM clusters are embedded in the ferromagnetic matrix in $x > 0.25$ samples. The latter coexistence induces considerable competition between FM and AFM exchange interactions and causes the above noted evolution of MR vs. $T$ dependences.

The results of neutron and electron diffraction presented in Refs. [8,14,15] (cited by G. Li et al in order to support their speculation on CO clusters embedded within the FM matrix of LCMO samples with $x > 0.25$) directly and unambiguously evidence that such CO (polaron-like) correlations are really observed in LCMO ($x = 0.30, 0.33$), but mainly in paramagnetic ($T > T_C$) state. The intensity of considered dynamical excitations sharply decreases in FM state, being equal to zero at $T \approx T_C - 20$ K - see Fig. 4 from [14]. Very recent data [16] on neutron diffraction completed by Monte Carlo modeling also evidence on developing of FM and AFM/CO correlations in LCMO ($x = 0.33$) at $T > T_C$, as well as on rapid suppression of AFM/CO ones just below $T_C$. Moreover, it is shown in [8] (cited by G. Li et al) that above noted CO clusters are observed in LCMO at $T < T_C$, but for $x = 0.2$, i.e. also in strong contradiction to speculation by G. Li et al, but in full agreement with data of Refs. [7,9,10].

At the same time, it is widely accepted now that the increasing of MR values at low $T \ll T_C$ may originate from enhanced contribution of GB-like effects to conductivity of different polycrystalline manganites due to the suppression of low $T$ minima on $\rho$ vs. $T$ curves by external $H$ [12,17-20]. One can easily see that the same effect, discussed in Refs. [12,17-20], is also observed in Fig. 3 from [5] for $x \geq 0.3$ and account for enhancement of MR at $T \ll T_C$. Moreover, MR vs. $T$ dependences analogous to these ones presented in Fig. 4 from [5] were obtained yet previously - see Figs. 4 and 2 from Refs. [11, 20], respectively, and explained by GB effects. Let us note also that the "broaden MR profile", claimed by G. Li et al as a specific feature of their polycrystalline LCMO samples with $x > 0.25$, was observed for single crystalline-like bulk $La_{0.78}Ca_{0.22}Mn_{0.9}O_{2.94}$ compound when sensing current (charge carriers) crosses through the single grain boundary [21]. In contrast, such broadening is absent for MR measurements inside this bulk grain - compare Figs. 4 and 5 in [21]. In addition, it is theoretically shown in [22] that the carriers' motion through GB region in polycrystalline manganites accounts for a broad maximum in $\rho$ vs. $T$ curve at $T_{max} < T_C$ - such one, which is observed by G. Li et al for $x = 0.37$ and $0.39$ samples - Fig. 3 from [5]. Finally, data of Ref. [23] (cited by G. Li et al) evidence that the proper heat treatment (melt-processing) of LCMO ceramic ($x = 0.33$) transforms GB effects-induced 'two maxima' shape of $\rho$ vs. $T$ curve (resulting in broadening of MR 'profile') to standard 'one sharp maximum' shape (without any broadening).

Of course the hypothesis by G. Li et al on increasing of MR at $T \ll T_C$ in LCMO system with Ca-content higher than 25% due to the enhancing influence of CO clusters may not be disregarded in principle. But for convincing argumentation in favor of such non-trivial idea the following conditions must be fulfilled:

(a) The considered LCMO samples must be well characterized;

(b) All GB-originated effects leading to the same (as claimed by authors) increasing of MR at $T \ll T_C$ must be eliminated/minimized by proper heat treatment, careful control of grain size, cationic homogeneity etc.;

(c) Convincing evidences on the existence of CO clusters in LCMO at $T \ll T_C$ for $0.25 < x \leq 0.40$ and on notable increasing of its volume with $x$ must be presented by authors or extracted from literature.

One can easily see that the above noted criteria are not fulfilled in [5] at all.

In summary, G. Li et al [5] tried to give additional insight into magnetoresistance in $La_{1-x}Ca_xMnO_3$ ($0 \leq x < 0.4$) system, exploring ac susceptibility and dc conductive measurements on polycrystalline samples. It is shown in this comment that, firstly, these samples were badly characterized by authors due to their ignorance of well-established basic features (see, e.g., [3,4,7]) of the phase diagram of LCMO system. Secondly, the published yet data on neutron and electron diffraction in LCMO strongly disagree with authors' hypothesis (increasing concentration of short range charge ordered antiferromagnetic clusters embedded in ferromagnetic matrix at $T \ll T_C$ for LCMO with $x > 0.25$). So, using by G. Li et al Refs. [8,14-15] for supporting of such speculation is absolutely groundless. Moreover, thirdly, the main result by G. Li et al (broadening of MR vs. $T$ 'profiles') may be easily and naturally explained taking into account enhanced influence of grain-boundary effects on low temperature conductivity in various manganites: polycrystalline samples [12,17-20], bicrystal epitaxial films [13] and bulk grains with single grain boundary [21]. Thus, all the above noted points allow us to conclude that the interpretation of the results presented in the paper [5] commented upon is very questionable.